# On the Dependence of Electromagnetic Phenomena on the Relativity of Simultaneity

DOUGLAS M. SNYDER
LOS ANGELES, CALIFORNIA

ABSTRACT

Maxwell's equations hold in inertial reference frames in uniform translational motion relative to one another. In conjunction with the Lorentz coordinate transformation equations, the transformation equations for the electric and magnetic field components in these reference frames can be derived. As the derivation of the Lorentz coordinate transformation equations depends on the relativity of simultaneity, and indeed on the argument on the relativity of simultaneity, electromagnetic phenomena indicate that human cognition is involved in the structure and functioning of the physical world.

TEXT

It is known that the relativity of simultaneity underlies the Lorentz coordinate transformation equations for two inertial reference frames in uniform translational motion relative to one another. It has been shown that an arbitrary decision on the part of the individual considering the two inertial reference frames as to which is the "stationary" and which the "moving" reference frame is involved in arguing the relativity of simultaneity. This arbitrary decision leads to the result that cognition is involved in the relativity of simultaneity and therefore in the structure and functioning of the physical world described with the use of the Lorentz coordinate transformation equations (Snyder, 1994).

The consideration of electromagnetic phenomena in terms of the special theory is particularly important because the special theory allows for a clear explanation of these common phenomena and, in contrast, explanations based in Newtonian mechanics do not. The arbitrary decision regarding the direction in which the relativity of simultaneity is argued is at the core of electromagnetic phenomena. This last result follows from the fact that the Lorentz coordinate transformation equations allow for the correct determination of electric and magnetic field components for inertial reference frames in uniform translational motion relative to one another. As Einstein wrote (1910/1993) in a quote that will be given later, one can hold that Maxwell's laws of electromagnetism are valid in two such inertial frames and use the Lorentz coordinate transformation to deduce the electric and magnetic field components in one of the reference frames once these field components are specified in the other reference frame. Thus, the relative character of forces due to electric and magnetic fields that intrigued Einstein are dependent on the argument on the relativity of simultaneity because this argument underlies the Lorentz coordinate transformation equations. This paper will attempt to demonstrate this point in some detail.

### THE RELATIVE CHARACTER OF FORCES DUE TO ELECTRIC AND MAGNETIC FIELDS

Allow that the electrically charged test particle is at rest in one of the inertial reference frames and that both electric and magnetic fields are present. In the inertial reference frame where the test particle is at rest, only a force associated with an electric field is exerted on the test particle. In the inertial reference frame where the particle is moving in a uniform translational manner, both the magnetic and the electric fields in general affect the test particle.

In the words with which Einstein began his first paper proposing the special theory of relativity:

> It is known that Maxwell's electrodynamics--as usually understood at the present time--when applied to moving bodies, leads to asymmetries which do not appear to be inherent in the phenomena. Take, for example, the reciprocal electrodynamic action of a magnet and a conductor [corresponding to the electrically charged test particle]. The observable phenomenon here depends only on the relative motion of the conductor and the magnet, whereas the customary view draws a sharp distinction between the two cases in which either the one or the other of these bodies is in motion. For if the magnet is in motion and the conductor is at rest, there arises in the neighbourhood of the magnet an electric field with a certain definite energy, producing a current at the places where parts of the conductor are situated. But if the magnet is stationary and the conductor is in motion, *no* electric field arises in the neighbourhood of the magnet. In the conductor, however, we find an electromotive force [due to the motion of the conductor in the magnetic field associated with the magnet], to which in itself there is no corresponding energy, but which gives rise--

> assuming equality of relative motion in the two cases discussed--to electric currents of the same path and intensity as those produced by the electric forces in the former case. (Einstein, 1905/1952, p. 37)

Considerations of the type just noted in the quote from Einstein were central to his development of the special theory of relativity. In a statement prepared for a meeting of the Cleveland Physics Society in 1952 honoring the centenary of Michelson's birth, Einstein wrote:

> What led me more or less directly to the special theory of relativity was the conviction that the electromotive force acting on a body in motion in a magnetic field was nothing else but an electric field. But I was also guided by the result of the Fizeau-experiment and the phenomenon of aberration. (Shankland, 1964, p. 35)

*The Lorentz Force Law*

To provide a bit of context for the quote immediately above, consider the following. The basic outline of the program in Newtonian mechanics is that a general law relates an external force applied to an object to the motion of this object, specifically that the object accelerates in direct proportion to the force applied to the object in the direction in which the force is applied. There are, in addition, various laws that specify the different forces, Newton's law of gravitational force being a prominent example of such a law. In Newtonian mechanics, the general force law is $F = ma$, where F is the external force applied to an object, m is the object's mass, and a is the acceleration associated with the application of the force. For electromagnetic phenomena, the basic specification of the force on an electrically charged particle is given by the Lorentz Force Law:

$$F = (qE) + (B \times v)$$

where F is the force experienced by the particle, q is the charge of the particle, E is the electric field, B is the magnetic field, and v is the uniform translational velocity, if any, of the electrically charged particle (Halliday & Resnick, 1960/1978). The term qE indicates that the particle experiences an electric force, irrespective of the particle's motion. The term B x v indicates that if the particle is moving and there is a magnetic field, the particle experiences a force orthogonal to the direction of the field and its motion. (The exception is where the motion of the particle and the direction of the field are in the same, or opposite, directions.) Thus in an inertial reference frame where the particle is moving in uniform translational motion through an electric field and a magnetic field, the particle experiences a force with two components, one due to the electric field and another associated with the magnetic field. If one considers the force exerted on the particle from the perspective of an inertial frame moving in the same manner as the particle, the particle experiences a force which is due only to an electric field. Irrespective of the inertial reference frame from which the particle is considered, as Einstein noted, "The observable phenomenon here depends only on the relative motion of the conductor [in our case, the electrically charged particle] and the magnet" (p. 37), not on the inertial reference frame from which it is considered.

<div style="text-align:center">

A MORE DETAILED ANALYSIS OF
THE RELATIVE CHARACTER OF
ELECTRIC AND MAGNETIC FIELDS THEMSELVES

</div>

Now here is Einstein's quote, alluded to earlier, in which he noted the dependence of the transformation equations for the field components in Maxwell's laws on the Lorentz coordinate transformation equations. He noted explicitly the consequences of this dependence.

> Let us apply the [Lorentz] transformation equations...to the Maxwell-Lorentz equations representing the magnetic field [and electromagnetic phenomena in general]. Let $E_x$, $E_y$, $E_z$ be the vector components of the electric field, and $M_x$, $M_y$, $M_z$ the components of the magnetic field, with respect to the system S [where the system S' is in uniform translational motion relative to S along the x and x' axes]. Calculation shows that the transformed [Maxwell-Lorentz] equations will be of the same form as the original ones if one sets
>
> Ex' = Ex
>
> Ey' = ß(Ey - v/c Mz)
>
> Ez' = ß(Ez + v/c My)
>
> Mx' = Mx
>
> My' = ß(My + v/c Ez)
>
> Mz' = ß(Mz - v/c Ey)
>
> [where $ß = (1/(1 - v^2/c^2)^{1/2}$ ].

# On the Dependence

The vectors ($E_x'$, $E_y'$, $E_z'$) and ($M_x'$, $M_y'$, $M_z'$) play the same role in the [Maxwell-Lorentz] equations referred to S' as the vectors ($E_x$, $E_y$, $E_z$) and ($M_x$, $M_y$, $M_z$) play in the equations referred to S. Hence the important result:

*The existence of the electric field, as well as that of the magnetic field, depends on the state of motion of the coordinate system.*

The transformed [Maxwell-Lorentz] equations [for inertial reference frames in uniform translational motion relative to one another] permit us to know an electromagnetic field with respect to any arbitrary system in nonaccelerated motion <u>S'</u> if the field is known relative to another system <u>S</u> of the same type.

These transformations would be impossible if the state of motion of the coordinate system played no role in the definition of the [electric and magnetic field] vectors. This we will recognize at once if we consider the definition of the electric field strength: the magnitude, directions, and orientation of the field strength at a given point are determined by the ponderomotive force exerted by the field on the unit quantity of electricity [the charged particle], which is assumed to be concentrated in the point considered and *at rest with respect to the system of axes.*

The [Lorentz] transformation equations [used to transform the Maxwell-Lorentz equations for electromagnetism] demonstrate that the difficulties we have encountered...regarding the phenomena caused by the relative motions of a closed [electric] circuit and a magnetic pole [associated with a magnetic field] have been completely averted in the new theory.

For let us consider an electric charge moving uniformly with respect to a magnetic pole. We may observe this phenomenon either from a system of axes S linked with the magnet [where B x v results in a force but no electric field], or from a system of axes S' linked with the electric charge [where a changing magnetic field generates an electric field]. With respect to S there exists only a magnetic field ($M_x$, $M_y$, $M_z$), but not any electric field. In contrast, with respect to S' there exists--as can be seen from the expression for $E'_y$ and $E'_z$--an electric field

On the Dependence

that acts on the electric charge at rest relative to S'. Thus, the manner of considering the phenomena varies with the state of motion of the reference system: all depends on the point of view, but in this case these changes in the point of view play no essential role and do not correspond to anything that one could objectify, which was not the case when these changes were being attributed to changes of state of a medium filling all of space. (Einstein, 1910/1993, pp. 140-141)

It is useful to provide the set of transformation equations for the electric and magnetic field components for Einstein's example where there is only a magnetic field in S where an electric charge is at rest. According to the electric and magnetic field transformation equations given by Einstein, the result for the scenario just described results in the following equations:

$E_y' = ß(- v/c\ M_z)$

$E_z' = ß(v/c\ M_y)$

$M_x' = M_x$

$M_y' = ß(M_y)$

$M_z' = ß(M_z)\ \cdot$

Thus, where no electric field exists in S and no electric force is thus exerted on the electric charge, in S' there are forces associated with both electric and magnetic fields on the electric charge.

It should be noted that where S' is the "stationary" frame and S the "moving" frame, for an electric charge at rest in S', a similar situation exists except for certain changes in certain electric field components in S due to the change in direction of the velocity of the reference frames relative to one another.

$E_y = ß(v/c\ M_z')$

$E_z = ß(- v/c\ M_y')$

$M_x = M_x'$

$M_y = ß(M_y')$

$M_z = ß(M_z')\ \cdot$

# On the Dependence

The similarity between this set of equations and the set of equations when S is the "stationary" inertial reference frame is due to the ability to consider either S or S' the "stationary" reference frame and the other frame the "moving" reference frame in arguing the relativity of simultaneity and employing particular set of the Lorentz coordinate transformation equations dependent on a particular direction in which the relativity of simultaneity is argued.

In sum, that Maxwell's equations hold in inertial reference frames supports the special theory, in particular the Lorentz coordinate transformation equations that allow for the derivation of the transformation equations for the electric and magnetic field components in these reference frames. As the derivation of the Lorentz coordinate transformation equations depends on the relativity of simultaneity, and indeed on the argument on the relativity of simultaneity, electromagnetic phenomena indicate that human cognition is involved in the structure and functioning of the physical world. One need go no further than to note that the integrity of the special theory depends on the ability to argue the relativity of simultaneity with either one of two inertial reference frames in uniform translational motion relative to one another the reference frame in which simultaneity is first defined in the argument. If this were not the case, then the fundamental tenet of the special theory that inertial reference frames are equivalent for the description of physical phenomena would not hold.